\documentstyle[11pt,newpasp,twoside]{article}
\def\edcomment#1{\iffalse\marginpar{\raggedright\sl#1\/}\else\relax\fi}
\reversemarginpar

\begin{document}
\title{Beyond M: Ultracool dwarfs in the Solar Neighbourhood}
\author {I. Neill Reid}
\affil{ Dept. of Astronomy \& Physics, University of Pennsylvania,
209 South 33rd Street, Philadelphia, PA 19104}

\begin{abstract}

I review the optical and near-infrared 
spectral characteristics of the extremely cool dwarfs grouped under
the new classification of type L. These include both very low-mass stars and
brown dwarfs, and we discuss the likely temperature
range. Finally, we consider the frequency
of binaries amongst very low-mass dwarfs, and the implications for 
formation mechanisms.

\end{abstract}

\section {Introduction}

One of the major goals of stellar astronomy in latter half of
the 20$^{th}$ century was the extension of
the Solar Neighbourhood census to encompass starlike objects at and
below the hydrogen-burning mass limit. That goal was realised in the 
past half-decade and, with the increased sensitivity provided by the new generation of
near-infrared sky surveys, we now have a collection of almost 200 dwarfs whose
spectral energy distributions indicate temperatures substantially
cooler than those of classic late-type M dwarfs, such as VB 8, VB 10 and LHS 2924. 
Determining the physical properties of these objects remains a high priority, not
only for the insight gained on the star formation process and the properties
of atmospheres at such low temperatures, but also as a bridge towards understanding 
the plethora of super-Jovian planets currently being revealed through radial velocity
observations of solar-type stars.

Davy Kirkpatrick's presentation at this meeting has already described the sequence of major
discoveries and the likely space density of very low-mass (VLM) dwarfs in the
Galactic disk. A full discussion of the background to this subject is
given by Reid \& Hawley (2000). 
This review concentrates on two main issues: the general characteristics of the ultracool
dwarfs now grouped under spectral class L, 
and recent results on low-mass binaries in the Hyades cluster and the 
field.  Adam Burgasser extends the discussion to include the even cooler T dwarfs.
In the complementary reviews in this session, John Stauffer covers recent results 
concerning very low-mass stars and brown dwarfs in the nearer open clusters and
star forming regions, while Peter Hauschildt outlines the progress made in
theoretical modelling of these very complicated atmospheres. 

\section {Brown dwarfs and low-mass stars}

The fundamental difference between low-mass stars and brown dwarfs is illustrated 
in figure 1, which plots the evolution of the central temperature of VLM 
dwarfs with masses
between 0.06 and 0.11 M$_\odot$. \footnote{ These predictions are taken from the theoretical models
computed by the Tucson group (Burrows {\sl et al.}, 1994; 1997); qualitatively similar results
are given by the Lyon models by Chabrier \& Baraffe, which are coupled with Allard \& Hauschildt's
model atmospheres (see Hauschildt's review, this conference).}
During the initial contraction
phase, the central temperature increases, regardless of mass, as potential energy is
transformed to heat. As the temperature rises above $\sim 3 \times 10^6$K, the initial
reactions of the PPI chain become possible, and hydrogen fusion is established.
The energy generated in the core leads to hydrostatic equilibrium, and the low-mass star
settles on the main-sequence, where it will continue burning hydrogen for 10$^{12}$ years or
more. 

\begin{figure}
\plotfiddle{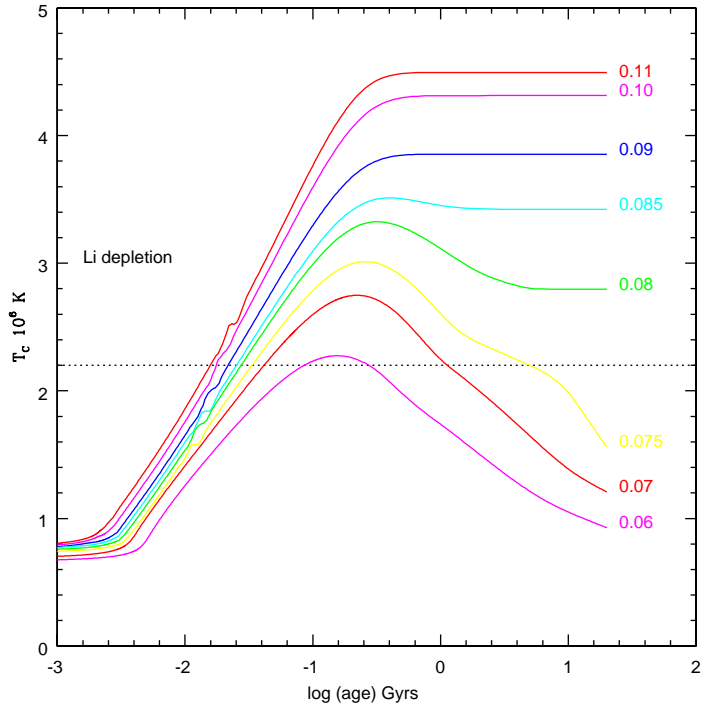}{9 truecm}{0}{50}{50}{-180}{-90}
\caption{The evolution of the central temperature in low-mass dwarfs}
\end{figure}

Low-mass dwarfs, however, are partially degenerate, and degeneracy absorbs some of
the energy generated both by contraction and the initial fusion reactions. This leads
to the downturn in temperature evident at masses below 0.09M$_\odot$. The
higher degeneracy in lower-mass objects can lead to T$_C$ falling below the critical 
value for maintaining hydrogen fusion. Robbed of the central energy source, the 
surface temperature and luminosity decrease steadily with time as the brown dwarf follows
a degenerate cooling curve. `Transition
objects' on the boundary between stars and brown dwarfs (0.075M$_\odot$ in these models; 
the Lyon models predict slightly lower masses) sustain fusion for several Gigayears before
fading into oblivion.

Two further characteristics require mention: first, VLM dwarfs have radii which are set by
degeneracy rather than gas pressure. As a result, all dwarfs with $M < 0.12 M_\odot$ have
radii within 15\% of the radius of Jupiter - the minimum lies at $\sim0.07 M_\odot$.
Second, the horizontal dotted line plotted on figure 1 marks T$_{Li}$, the temperature required for
fusion of lithium (by reaction 2 in the PP chain). Late-type dwarfs are fully convective, so
if $T_C > T_{Li}$, primordial lithium is depleted; conversely, if T$_C$ fails to exceed the
critical value for sufficient time, lithium is at most partially destroyed. This is the
underlying basis for the lithium test devised by Rebolo {\sl et al.} (1992): dwarfs
later than spectral type M6 with Li 6708\AA\ absorption can be identified as brown dwarfs
with masses below $\sim0.06 M_\odot$.

\begin{figure}
\plotfiddle{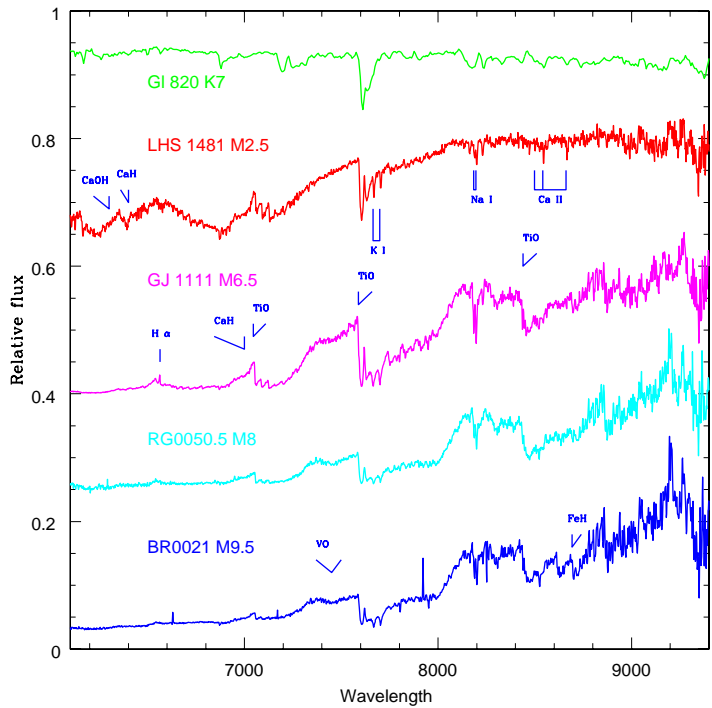}{9 truecm}{0}{50}{50}{-180}{-90}
\caption{ Far red spectra of late-type M dwarfs}
\end{figure}

\section {Spectral class L}

Spectral class M is characterised by the presence of strong absorption bands due to the diatomic
molecule titanium oxide, TiO (Morgan, Keenan \& Kellman, 1943). Figure 2 shows far-red spectra
of dwarfs spanning spectral classes K7 to M9.5. The most prominent features are marked, including
CaOH and CaH in mid-type dwarfs and VO and FeH absorption in the later-type dwarfs. H$\alpha$ is
also evident in emission in RG0050.5 and GJ 1111.

\begin{figure}
\plotfiddle{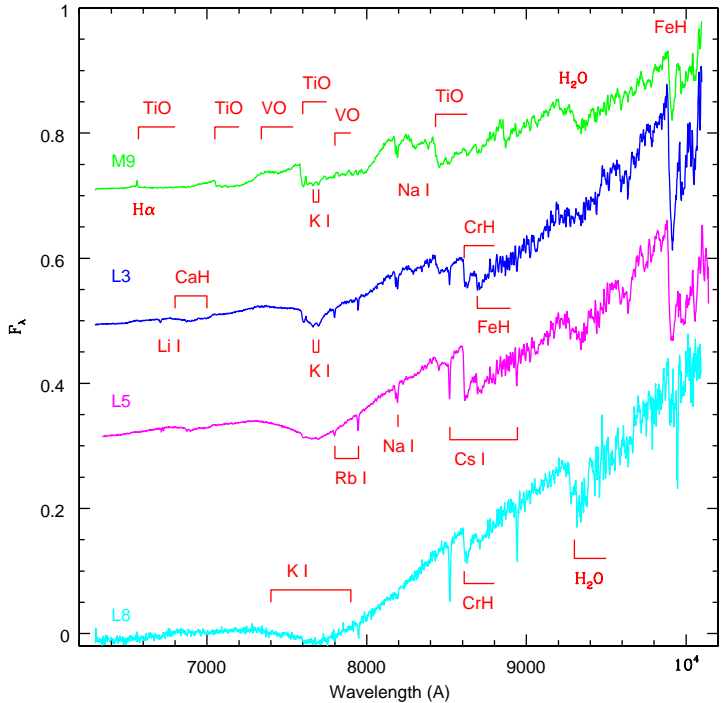}{9 truecm}{0}{50}{50}{-180}{-90}
\caption{ Spectral class L: LRIS spectra of typical examples }
\end{figure}

Figure 3 plots a skeletal sequence for spectral class L, following the classification scheme
outlined by Kirkpatrick {\sl et al.} (1999 - K99). The most important characteristics 
are the decreasing strengths of TiO and VO absorption as the overall
energy distribution steepens from blue to red. The main molecular features are 
metal hydrides, CaH, FeH and CrH (and MgH at shorter wavelengths), and H$_2$O: CaH peaks
in strength in the earlier L dwarfs, FeH at mid-L and CrH and H$_2$O grow in strength from
early to late L. The only other prominent features are strong atomic lines of the alkali
elements, notably Na, K, Rb, Cs and (in some cases) Li, the presence of the last 
identifying some ($\sim$25\%) L dwarfs as lower-mass brown dwarfs. Note 
the substantial broadening of the 7666/7699 \AA\ KI resonance doublet in later L dwarfs,
reaching a full-width of
$\sim1000$\AA\ by spectral class L8 (the Na I D lines, another resonance doublet, reach similar
strengths by L5, Reid et al, 2000a).

The explanation for this behaviour is generally agreed to lie with the formation
of atmospheric dust particles as the dwarf cools to temperatures below $\sim2600$K
(spectral type$\approx$M6). Originally suggested by Tsuji {\sl et al.} (1996), partly as
a means of explaining the relatively shallow near-infrared H$_2$O absorption in late-type
M dwarfs (see further below), this mechanism has been investigated by several groups
(Fegley \& Lodders, 1996; Lodders, 1999; Burrows \& Sharp, 1999; Hauschildt {\sl et al.}, 
this conference). In brief, the transformation of TiO and VO from gas to solid phase through the formation of
materials like perovskite (calcium titanate,  CaTiO$_3$), solid VO and
silicates, such as enstatite ( Mg$_2$SiO$_3$) and fosterite (Mg$_2$SiO$_4$), leads to
both weaker molecular absorption, and a substantial reduction in the opacity
at optical wavelengths. That, in turn, leads to the increasingly smooth energy distribution
at later spectral types, while the high atmospheric transparency places the $\tau=1$ `photosphere' at
substantial physical depths, leading to high column densities for the alkali elements and strong lines. 
The continued presence of metal hydrides is
analogous to the situation in cool, metal-poor extreme subdwarfs: few metallic atoms, but lots of H.

\begin{figure}
\plotfiddle{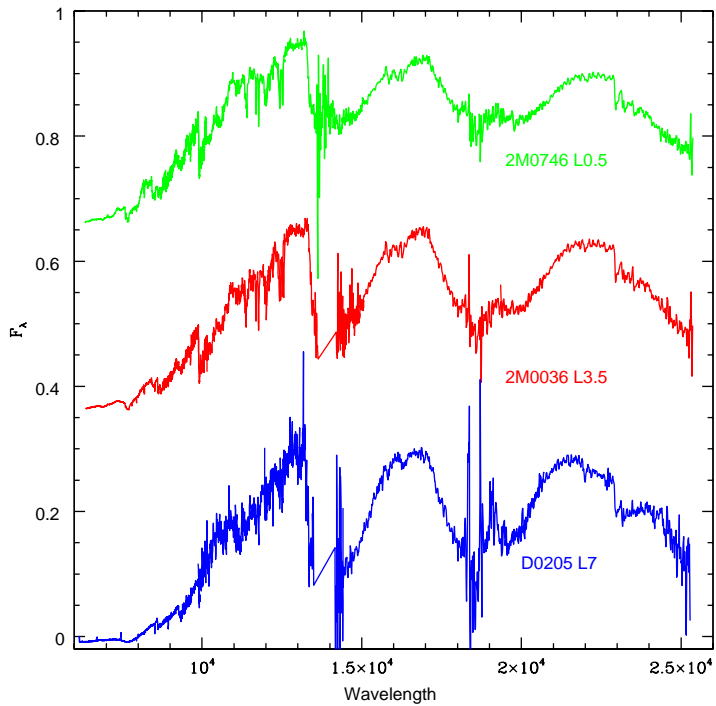}{9 truecm}{0}{50}{50}{-180}{-90}
\caption{ Near-infrared spectra of L dwarfs (CGS4 data - Reid {\sl et al.}, in prep.) }
\end{figure}

L dwarfs are faint at optical wavelengths, rendering photometry and
spectroscopy extremely difficult; this is the main reason why wholesale
discovery was postponed until the DENIS and 2MASS surveys came on line. Indeed, 
there little question that detailed observations would have
been impossible without Keck, HIRES (Vogt et al, 1994) and LRIS (Oke et al, 1995). 
Capitalising fully and efficiently on future discoveries demands extension of the 
optical spectral sequence to the near-infrared wavelengths where most of the
energy is produced. That transfer is now becoming possible with the development of
intermediate-resolution near-IR spectrographs. 

Figure 4 plots UKIRT CGS4 spectra of an early-, mid- and late-type L dwarf.
There is  greater similarity between spectral classes L and M at near-infrared
wavelengths: differences are more a matter of degree than of kind.
The strongest features are the H$_2$O bands at 1.4 and 1.8 $\mu$m, features also
present in the terrestial atmosphere, and the CO bandhead at 2.3 $\mu$m.
Pressure-induced H$_2$ absorption can also be present at $\sim2 \ \mu$m in
some late-type L dwarfs (Tokunaga \& Kobayashi, 1999).
Hydride bands, notably FeH, and atomic lines due to Na, K and other metals
are also present, primarily in the J band. Currently available data, which 
spans only a relatively small number of dwarfs, indicates that variations in the near-infrared
features are well correlated with the optical variations: that is, an independently
constructed near-infrared spectral sequence can be expected to produce the same
ordering as optical data (McLean {\sl et al.}, 2000; Leggett {\sl et al.}, in prep.; 
Reid {\sl et al.} in prep.). 

\section {On spectral classification}

The primary definition of subclasses for spectral class L rests with Kirkpatrick {\sl et al.}
(K99), who subdivided the  initial 2MASS detections into classes L0 to L8.
Mart\'in {\sl et al.} (1999b) have outlined an alternative subdivision which
differs in some respects from the K99 system, notably at later types (L7 $\approx$ L5 in this system).
I believe that the latter system is flawed at a fundamental level in its conception. 

Spectral classification is astronomical botany: it is an exercise in pure morphology, 
an ordering and organisation based solely on appearance. The hope and expectation is
that by undertaking this exercise, one is also producing a sequence whose order is
determined by some underlying physical parameter (usually temperature). 
That expectation can be fulfilled through a sensible choice of morphological criteria
- substantive variations rather than minutiae. 
However, one must carefully divorce the act of classification from theoretical interpretation:
classify, then interpret. The basis for this approach is obvious:
theoretical models are ephemeral, periodically replaced with revised and
improved analyses; the equivalent widths, line strengths and flux distribution of a 
given spectroscopic observation are, within observational uncertainty, 
fixed. A spectral classification system
which must change with the release of each  new set of theoretical models is of
little use to anyone.

Spectral classification is based on what an object {\sl looks like}, not what it {\sl is}.

Mart\'in {\sl et al.} violate this separation by tying their classification system
directly to temperatures estimated from a particular set of models. As discussed further 
below, there are also grounds for doubting the temperature scale adopted in
Mart\'in {\sl et al.'s} system, but that is a separate matter.

One should not expect the conflict between these two proposed systems
to be decided from above - classification
systems are decided by acclamation, rather than legislation.
To quote the originator of a well established system,
\begin{quote}
The MK system has no authority whatever; it has never been adopted as an official
system by the International Astronomical Union - or by any other astronomical
organisation. Its only authority lies in its usefulness; if it is not
useful, it should be abandoned. \emph{ W. W. Morgan (1979)}
\end{quote}
In the same way, the system chosen to subdivide spectral class L (and T) will
be decided informally, by use, rather than by mandate. 

\section {Brown dwarf weather}

In modelling stellar atmospheres, one usually thinks of the photosphere, the
$\tau=1$ level, as a surface-like layer at a particular physical depth with a 
particular temperature.
This approximation is reasonable for solar-type stars, but starts to break down
in M dwarfs, where the feature-rich spectrum can lead to our sampling 
very different depths at different wavelengths. The situation for cool brown
dwarfs is likely to be even more complex: not only does the formation and atmospheric
dispersal of dust affect the evolution of the energy distribution, but topographical
variations (bands and zones {\sl a la} Jupiter, or clouds) may well lead to
temporal variations (i.e. weather). Thus modelling brown dwarf atmospheres may
well be more akin to planetary atmosphere analysis than classical stellar studies
(replace photosphere with tropopause).

\begin{figure}
\plotfiddle{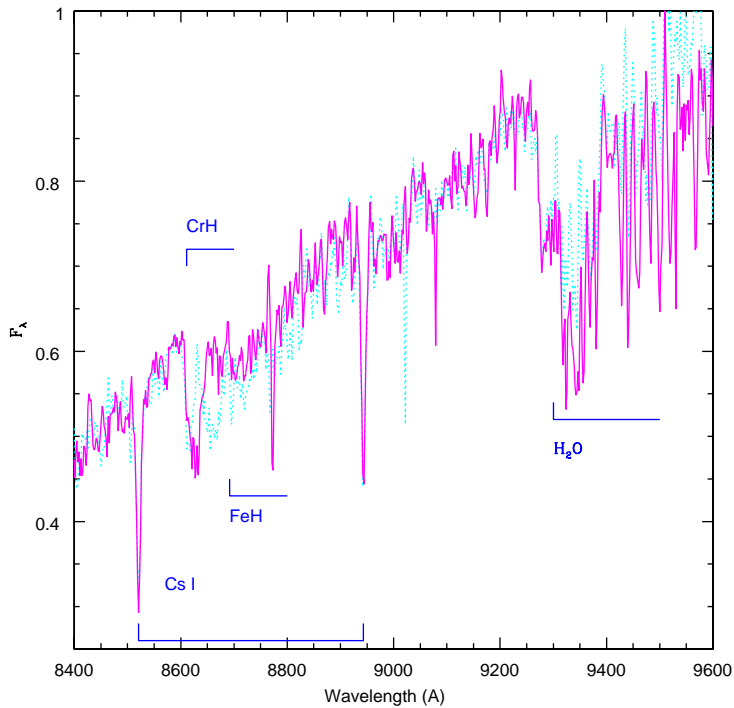}{9 truecm}{0}{50}{50}{-180}{-90}
\caption{ Temporal variations in the spectrum of Gl 584C: the
dotted line plots observations on 24/12/98; the solid line, data
from 25/12/98 (Kirkpatrick et al, in prep.) }
\end{figure}

Some late-type M dwarfs are known occasionally to exhibit photometric
variations, usually assumed to be associated with magnetic activity in the
form of spots (eg Krishnamurthi {\sl et al.}, 1998; Reid \& Hawley, 2000, ch. 5).
Most attempts to detect such variations in ultracool dwarfs, notably by
Tinney \& Tolley (1999), have been unsuccessful. However, examples have
recently come to light: first, Bailer-Jones \& Mundt (2000) have measured cyclic
photometric variability at the $\pm0.05$ mag. level in I-band monitoring of the
L1.5 dwarf, 2MASSW J1145572+231730. 2M1145 is one of the relatively few L
dwarfs with H$\alpha$ emission (see Gizis {\sl et al.}, 2000, for a more
extensive discussion of chromospheric activity in ultracool dwarfs), suggesting
that the photometric variations might be explained by the usual rotation-plus-spots model. 
The data are well-phased to a period of 7.12 hours which,
interpreted as rotation, implies an equatorial velocity of $\sim 17$ kms$^{-1}$ for 
$R = 0.1 R_\odot = 1 R_J$. 

More unusual variations have been detected in the late-type L
dwarf, 2MASSW J1523226+301456, spectral type L8. 
This dwarf has strong lithium absorption, and is a common proper-motion
companion of the G dwarf binary, Gl 584AB, lying at a separation of 3600 AU 
and a distance of 18.6 parsecs from the Sun. The association not only allows us to
identify Gl 584C as one of the lowest luminosity brown dwarfs known, with M$_J$=15.0,
0.5 magnitudes brighter than Gl 229B, but also permits estimation of the age (from
chromospheric activity in the G dwarfs) and hence a mass   of 
$\sim0.045 M_\odot$ (Kirkpatrick {\sl et al.}, in prep.). We
observed 2M1523 on December 24 and 25, 1998, and figure 5 superimposes (with
no adjustments) the 
independently flux-calibrated spectra. It is clear that while there is 
excellent agreement in the overall shape of the pseudo-continuum and the
Cs I absorption lines, there are significant variations in the strengths of 
molecular features. In particular, the CrH/FeH region at 8650\AA\ was filled in
to some extent on the 24th, while the 9300\AA\ water band is weaker. 
Subsequent observations (in March and July, 1999) show no further significant
changes with the spectrum closer to our observations on Dec 24th.

The simplest interpretation of these results, driven mainly by the observed variation in
H$_2$O strength,  is that the effective temperature on
the 25th was lower than the average. The conventional interpretation would be
to scribe this variation to the presence of a giant star spot. However, 2M1523 shows
no evidence for significant chromospheric activity (${L_\alpha \over L_{bol}} < 10^{-6}$).
An alternative possibility is that we detected the presence of a substantial cloud, which
moved the $\langle \tau \rangle = 1$ level to higher physical heights and lower temperatures.
These observations are difficult and time consuming, but
more extensive monitoring (an unpopular word for TACs)
of the latest type L dwarfs  might
provide insight into the detailed structure of these cool atmospheres.

\section {L dwarf temperatures and the L/T transition zone}

To the best of our ability to test such matters, 
ordering M and L dwarfs by spectral type produces a rank ordering by effective temperature.
Tying this relative scale to absolute values is important not only in
the analysis of individual objects, but in using the statistical
results from wide-field surveys to set constraints on the initial mass function (Kirkpatrick, 
this conference; Reid {\sl et al.}, 1999). Full model atmosphere analyses of these
complex spectra remain extremely difficult (Hauschildt, this meeting), but, in principle, 
one might hope to use the appearance and disappearance of individual features to set
constraints on T$_{eff}$ (with the over-riding caveat that `effective temperature' may
have less conceptual validity in these objects). Thus, VO is predicted to solidify at
a temperature between 1700 and 1900K. Observations show that VO disappears from the
spectrum at $\approx$L4 (K99, Burrows \& Sharp, 1999). 

The most significant change in the spectral energy distribution of cool brown 
dwarfs comes at near-infrared wavelengths, with the change from CO to CH$_4$
as the dominant atmospheric repository of carbon. This is accompanied by the
onset of strong overtone CH$_4$ absorption at 1.6 and 2.1 $\mu$m: 70\% of the
flux emitted at H and K in L dwarfs is absorbed, but J is little affected, leading to
(J-K) colours closer to A-type dwarfs than L dwarfs. This effect was originally
predicted by Tsuji (1964), but the discovery of Gl 229B still came as 
something of a surprise to the astronomical community. The appearance of these
near-infrared methane bands marks the transition from spectral type L to T (K99)
\footnote{Spectral type T is defined specifically by the presence of overtone CH$_4$ absorption
in the H and K passbands. Methane is also present in L dwarf atmospheres, albeit
well above the `photosphere', as evidenced by
the recent detection of the 3.3$\mu$m fundamental band by Noll {\sl et al.} (2000). This
fact has been used to argue against the use of the term `methane dwarf'; on the other hand,
carbon is present in more than carbon stars.}.

\begin{figure}
\plotfiddle{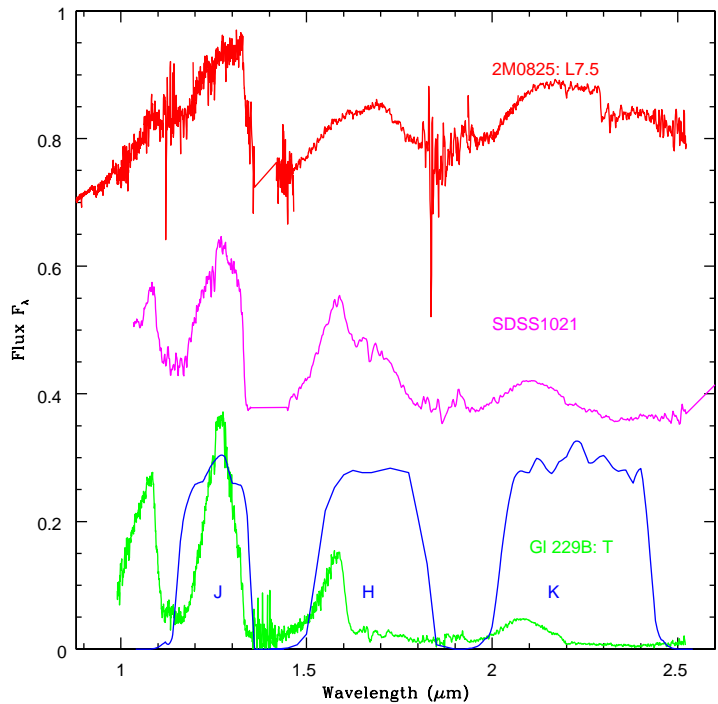}{9 truecm}{0}{50}{50}{-180}{-90}
\caption{ The transition from L to T: data for SDSS1021 courtesy of S. Leggett; 
data for Gl 229B from Geballe et al, 1996 }
\end{figure}

Our prospects of understanding brown dwarf characteristics over the L/T transition
has received a boost through
the recent discovery of several early-type T dwarfs using data
obtained by  the Sloan Digital Sky Survey (Leggett {\sl et al.}, 2000). As figure 6
illustrates, the near-infrared methane bands in these objects are less
saturated than in the `classic' Gl 229B-like T dwarfs. In addition, the H$_2$O
band at $\sim1.1 \ \mu$m, which is just starting to appear in the L7.5
dwarf, 2M0825, is not as strong as in Gl 229B. 
The effect of these spectral changes on the near-infrared colours is shown in figure 7.
While the early-type T dwarfs have extremely red (I-J) colours, their
JHK colours are similar to late-K/early-M giants, a
fact which makes their discovery based on near-infrared data alone an extremely
unlikely eventuality.  Combination of Sloan's deep, far-red data
with 2MASS JHK (particularly J) photometry will provide an extremely effective
method of locating these objects (see Kirkpatrick, this conference).

\begin{figure}
\plotfiddle{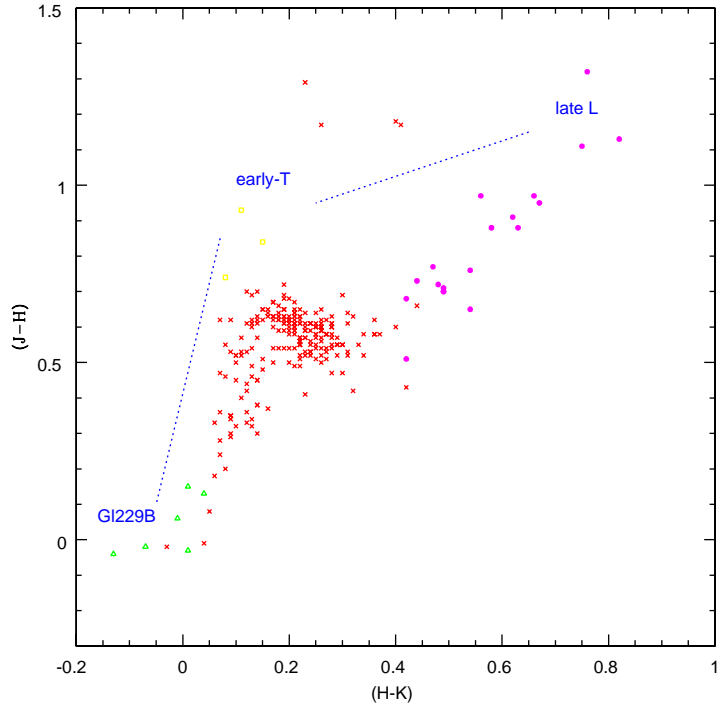}{9 truecm}{0}{50}{50}{-180}{-90}
\caption{ Brown dwarf L/T evolution in the JHK diagram: crosses are
nearby stars;  solid points
are L dwarfs; open triangles, Gl 229B-like T dwarfs; and open squares,
early T dwarfs from Leggett {\sl et al.} (2000).  }
\end{figure}

Two attempts have been made to tie L dwarf spectral types to effective temperatures.
Kirkpatrick {\sl et al.} (K99, 2000) have used the appearance/disappearance of 
individual spectral features as the basis for a scale running from T$_{eff} \approx 2000/2100$K
at spectral type L0 to $\approx 1300/1400$K for L8 dwarfs, such as 2M1523. For reference, 
the effective temperature of Gl 229B has been determined as $960\pm70$K (Marley {\sl et al.}, 
1996), and some of the more recently discovered T dwarfs may be significantly hotter
(Burgasser, this conference). This scale therefore envisages a relatively small temperature
difference between the SDSS early T-dwarfs and the latest L dwarfs. 

In contrast, Basri {\sl et al.}(1999) have used detailed line-profile analyses in
conjunction with the Allard \& Hauschildt atmosphere models to derive a much hotter
temperature scale. To give a specific example, they estimate an effective temperature
of $\sim1750$K for the L7\footnote{For the reasons given in section 4, we eschew the
Mart\'in {\sl et al.} classification scheme and quote spectral types on the K99 system.}
dwarf DENIS-P J0205-1159, which would be assigned T$_{eff} \sim 1400$K on the K99 calibration.
Thus, this scheme envisages a difference of almost a factor of two in temperature between 2M1523 and
Gl 229B - and a substantial number of `missing' (early-type T?) brown dwarfs, with
consequent impact on $\Psi(M)$.

\begin{figure}
\plotfiddle{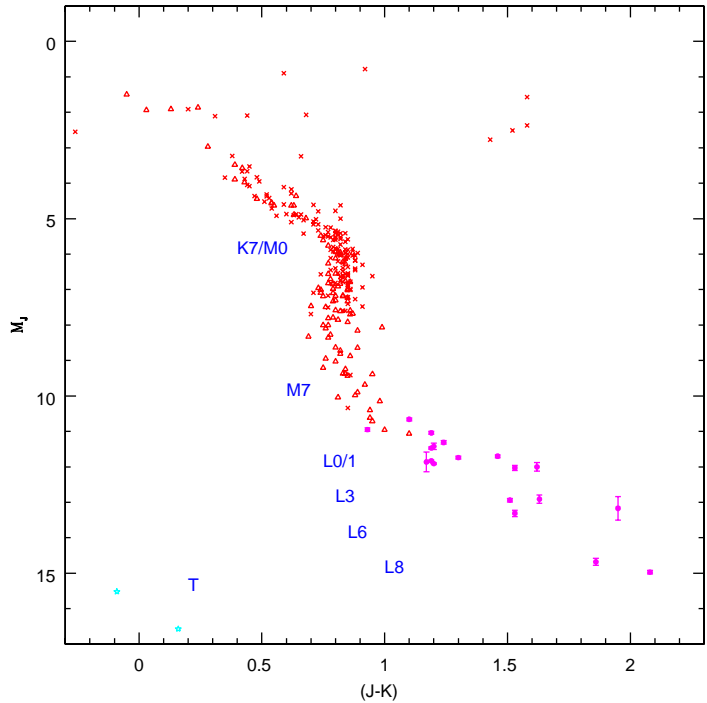}{9 truecm}{0}{50}{50}{-180}{-90}
\caption{ The near-infrared colour-magnitude diagram: triangles identify stars from the 
8-parsec sample with photometry by Leggett; crosses mark nearby stars with 2MASS
JHK$_S$ data; and solid points are late-M and L dwarfs with photometry from either
2MASS or USNO. }
\end{figure}

Several arguments, however, can be marshalled against the hotter scale:
\begin{itemize}
\item Our 2MASS follow-up observations now cover $\sim 4000$ sq. deg., or 10\% of the
sky, and we have identified no L dwarfs later than spectral type L8 ((J-K)$\approx 2$). 
Lower-temperature, lower-luminosity T dwarfs have been found within the same area. 
We conclude that L dwarfs later than L8 are extremely rare; indeed, L8 may mark
the end of the L dwarf sequence.

\item Evolution from spectral type L8 to T involves a substantial
change in (J-K) colour (figures 7 and 8); however, that need not imply a substantial
change in temperature. Theoretical calculations indicate that the transition
from CO-dominant to CH$_4$-dominant atmospheres spans only 30 to 40K (Lodders, 1999). 
Brown dwarf atmospheres are extremely complex, but this result does not appear consistent with
the large L/T temperature difference envisaged in the Basri {\sl et al.} scheme.

\begin{figure}
\plotfiddle{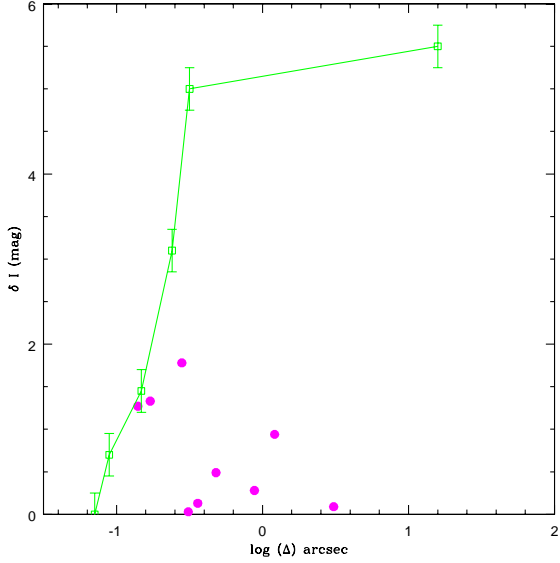}{7 truecm}{0}{40}{40}{-150}{-70}
\caption{Detection limits for HST imaging. The sensitivity curve plotted is for the
Planetary Camera at 8000\AA, but similar constraints apply to NICMOS imaging
at 1.1$\mu$m. Note that all of the detected Hyades binaries lie well above the magnitude
limit.}
\end{figure}

\item Figure 6 shows
that the J-band is relatively unaffected by the transition from late-L to T. One might
therefore expect M$_J$ to provide a reliable indication of M$_{bol}$. Indeed, Gl 229B has
M$_{bol} = 17.7$, or BC$_J = 2.2$; the latter value is within 0.2 magnitudes of
that measured for late-type M dwarfs. Since brown dwarfs are degenerate, we know that 2M1523 and
Gl 229B have radii which differ by less than 15\%. Thus, a $\Delta M_J = 0.5$ mag corresponds to
$\Delta M_{bol} < 1$ mag., or, from
\begin{displaymath}
{L_1 \over L_2} \ = \ ({R_1 \over R_2})^2 \ ({T_1 \over T_2})^4 
\end{displaymath}
a temperature difference of less than 25\%. Given T$_{eff} = 960$K for Gl 229B, this implies
T$_{eff} \sim 1250$K for 2M1523. \\
Conversely, we can invert the problem: if 2M1523 has an effective temperature of
1700K, then M$_{bol} \sim 15.6$, and BC$_J \sim 0.6$ magnitudes, radically smaller than
empirical estimates. 

\item Figure 8 plots the (M$_J$, (J-K)) colour-magnitude diagram defined by nearby stars
with well-determined trigonometric parallaxes. L0/L1 dwarfs have M$_J \sim 12$, while
2M1523, L8. has
M$_J$=15.0; as noted above, Gl 229B is only 0.5 magnitudes fainter. Basri 
{\sl et al.} bracket the former interval by $\delta T \sim 600$K (2300 to 1700K), 
and the latter by $\delta T \sim 700$K; the photometric scale gives $\delta T \sim 750$K
(2100 to 1350K) and $\sim350$K, respectively. 

\end{itemize}

Clearly, further observations ana analyses are required before we can arrive at
a definitive temperature scale for late-type dwarfs. Parallax measurements for 
early-type T dwarfs and a larger sample of late L dwarfs will be particularly useful.
However, in my view, the current balance of evidence favours the cooler scale.

\section {Low mass binaries}

Binary systems are important astrophysical tools. If full orbital solutions can be determined, 
individual components can be used to improve mass-luminosity(-age)
relations, while the frequency of binary systems, and the distribution of parameters
(mass ratios, semi-major axes, eccentricities, etc) as a function of systemic mass, can
set constraints on star formation mechanisms. The classic work on solar-type
stars is Duquennoy \& Mayor's (1991) survey of nearby stars, which found a binary fraction
of at least 60\%. Studies have consistently shown that binary frequency declines amongst
lower-mass stars: Fischer \& Marcy (1992) derive a multiplicity of $\sim35\%$ (i.e. $\approx$2
of 3 M dwarf systems are isolated, single stars), a result confirmed by Reid \& Gizis (1997a). 

\begin{figure}
\plotfiddle{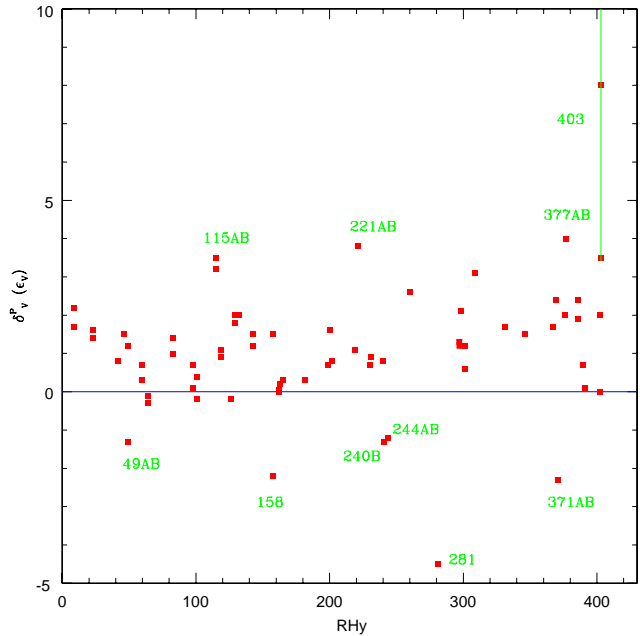}{8 truecm}{0}{45}{45}{-150}{-70}
\caption{ Velocity offset  [$\delta_V$ = (Obs - predicted)/$\epsilon_V$] from Keck HIRES observations 
of Hyades late-M dwarfs. The x-axis is the star number from Reid (1992), increasing with increasing RA.}
\end{figure}
Further data on the frequency of low mass binaries have been provided by several HST
imaging surveys, notably Planetary Camera observations of M dwarfs in the field (Reid \& Gizis, 1997a)
and Hyades cluster (Reid \& Gizis, 1997b), and NICMOS data for low-mass stars and
brown dwarfs in the
Pleiades cluster (Mart\'in {\sl et al.}, 2000). Those observations allow one to search for
resolved companions at separations of less than 0.2 arcseconds, with the sensitivity limits
extending well into the brown dwarf r\'egime for the younger dwarfs in the open clusters. 
The results derived for the two M dwarf samples (8/41 binary or multiple systems in the field; 9/53 in
the Hyades) are broadly consistent with expectations based on the nearby M dwarfs. The Pleiades
survey, on the other hand, revealed no resolved systems, a $\sim2\sigma$ discrepancy with
respect to the field. 

We have recently extended the Hyades survey by obtaining Keck HIRES data for all of 53
M dwarfs observed with HST (Reid \& Mahoney, 2000), using a variety of techniques to
search for close, spectroscopic binaries. Perhaps most interesting is the fact that we
can use single observations to identify candidate binaries: all of these stars are confirmed
as Hyades cluster members through both astrometry (proper motions) and chromospheric
activity; as a result, we can predict the radial velocity of individual
stars using the known cluster space motion and the angular distance from the convergent
point; stars which deviate significantly from the expected value are likely to 
be exhibiting binary motion. Figure 10 plots the results of this exercise, where
the convergent point is taken from Perryman {\sl et al.'s} (1998) Hipparcos analysis.
(The systematic trend in $\delta_V$ with RA suggests that either the adopted
cluster motion, V$ = 45.72$ kms$^{-1}$, or the convergent point requires revision.)

\begin{figure}
\plotfiddle{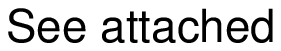}{10 truecm}{270}{38}{38}{-150}{260}
\caption{ An alternative model for brown dwarf binaries }
\end{figure}

All of the stars which deviate from the main body of data in figure 10 are confirmed
or highly-probably binaries. One star in particular stands out: RHy 403. At M$_V = 14.4$, this
is one of the lowest luminosity Hyades members. Following our initial discovery of significant
velocity variations,  we monitored this star over Dec 28/29 and 29/30 in 1999 (some may notice an
irritating  consistency in TAC allocations for these projects). The HIRES data show clear binary
motion, well-matched by a sinusoid with period 1.276 days, amplitude $\pm 40$ kms$^{-1}$.
The system is single-lined, setting an upper limit of $\sim0.095 M_\odot$ on the secondary, while
the mass function sets a lower limit of 0.06M$_\odot$. This is the best candidate discovered to date
for a brown dwarf member of the Hyades - and it is the {\sl only} candidate. The `brown dwarf desert'
suspected amongst G dwarfs (Marcy \& Butler, 1998) appears to extend to M dwarfs.

What about brown dwarf/brown dwarf binaries (figure 11)? Such objects have been known
for some time: two of the original three field L dwarfs discovered in the DENIS brown
dwarf mini-survey (Delfosse {\sl et al.}, 1997) and one of the early 2MASS
discoveries prove to be equal-luminosity (therefore
equal-mass) systems, with sub-arcsecond separations (Mart\'in {\sl et al.}, 1999a; 
Koerner {\sl  et al.}, 1999). Initial results suggested that such systems
may be more frequent than amongst M dwarfs. However, we have been using HST to
obtain F814W (I-band) images of a larger sample of L dwarfs, and those data
point to a different conclusion. Only four of the 20 L dwarfs imaged successfully 
are resolved (Reid {\sl et al.}, 2000b). All four have component separations of a few tenths of an 
arcsecond, corresponding  to $\Delta < 10$ AU (one system, 2M1146, was known
as binary prior to our observations).  Two systems (2M0920 and 2M0746) are at sufficiently 
small separations that orbital determinations, and dynamical mass estimates, may be
possible in a matter of a decade or so.  

\begin{figure}
\plotfiddle{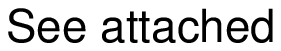}{8 truecm}{0}{45}{45}{-150}{-70}
\caption{The L dwarf binary, 2M0850+1057: HST F814W observations}
\end{figure}

Three of the four HST-observed L dwarfs have components with near-equal
luminosities.
Since these low-mass objects evolve rapidly with time, this
equality implies a comparable equality in mass, as with the 
previously known systems. One system, 
however, 2M0850+1057, has components which differ by 1.8 magnitudes at I
(figure 12). 2M0850AB is one of the latest type L dwarfs, spectral type L6, 
and has a well-determined trigonometric parallax from US Naval Observatory
astrometry (Dahn {\sl et al.}, 2000). 
Deconvolving the relative contributions
of the two components, it becomes clear that 2M0850B is extremely interesting,
with an absolute magnitude, M$_I$, between the lowest-luminosity L8 dwarf
(2M1523) and T dwarfs (figure 13).
 
The detection of lithium in 2M0850AB indicates a mass below 0.06M$_\odot$ (for both
components), with a mass ratio ${M_s \over M_p} \sim 0.8$.
The  inferred J magnitude for 2M0850B is M$_J = 15.2$, 0.3 magnitudes brighter than Gl 229B. 
Near-infrared spectroscopy of the (unresolved) 2M0850AB system shows no evidence
for CH$_4$ absorption in the K band, suggesting that 2M0850B (which contributes $\sim15\%$
of the flux at those wavelengths) is spectral type L. This system, and others
like it offer the prospect of bridging the gap from spectral type L to type T.

\begin{figure}
\plotfiddle{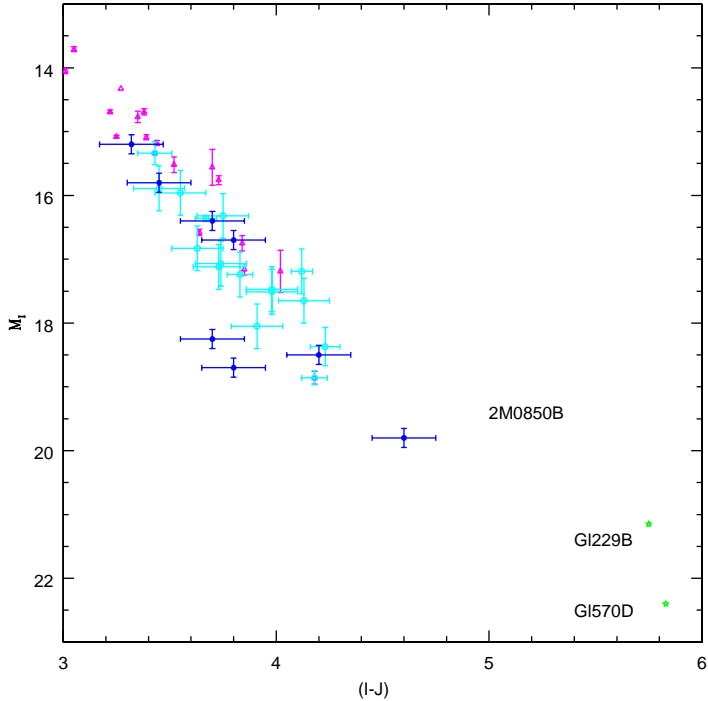}{9 truecm}{0}{50}{50}{-180}{-90}
\caption{ Newly-resolved L dwarf binaries (solid points) in the M$_I$, (I-J) plane}
\end{figure}

On the statistical front, our HST survey of field L dwarfs shows the same $2\sigma$
discrepancy of wide ($\Delta > 10$ AU) binary systems noted by Mart\'in {\sl et al.} (2000)
in their Pleiades survey. These two surveys target systems with similar masses: the
majority of the Pleiads are spectral type M, but the hotter temperatures reflect 
a younger age than the average in the field ($\sim125$ Myrs vs $\sim1$ Gyr). Wide binaries 
are rare amongst VLM, ultracool dwarfs. However, both L dwarfs and T dwarfs are found
as wide, common proper-motion companions of main-sequence stars: Gl 229B is 45 AU from 
the M0.5 dwarf, Gl 229A; GD 165B is 110 AU from the white dwarfs GD 165A; G 196-3B is 340 AU
from the M dwarf, G196-3A;  Gl 670D lies 1525 AU from the triple system, Gl 670ABC; and, most
extreme, Gl 584C is 3600 AU from the G-dwarf binary, Gl 584AB. Many of these systems
can be dated using the characteristics of the main-sequence companion, notably chromospheric
activity, allowing masses to be estimated for the brown dwarf companion.

Pulling these results together, 
\begin{enumerate}
\item VLM/VLM binaries are rare at separations exceeding 10 AU;
\item The preference for equal-mass systems at small $\Delta$ may be a selection effect, but
echoes a similar preference amongst M dwarfs (Fig. 9; Reid \& Gizis, 1997a,b); 
\item Wide binaries involving a more massive main-sequence star and a VLM dwarf are not uncommon.
\end{enumerate}

\begin{figure}
\plotfiddle{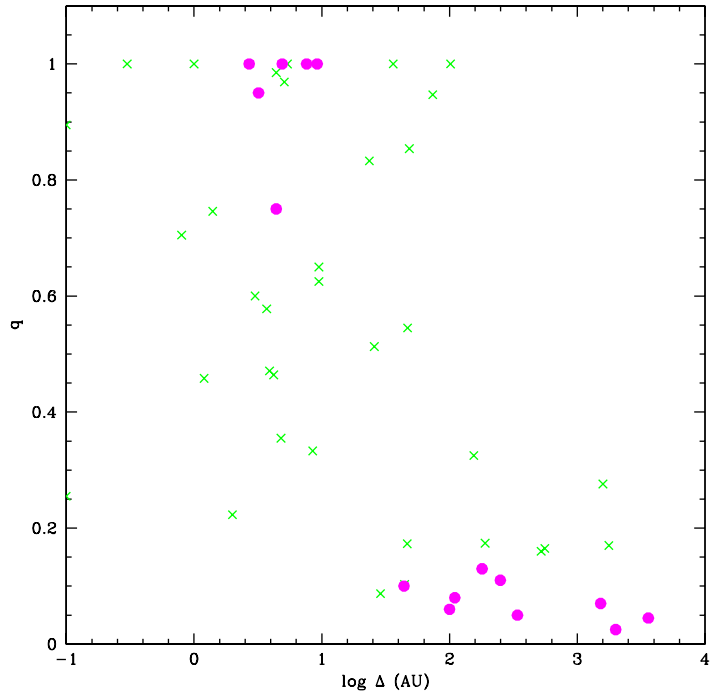}{9 truecm}{0}{50}{50}{-180}{-90}
\caption{ Binary mass ratios as a function of log(separation): solid points are systems
with at least one L- or T-dwarf component; crosses are systems with M-dwarf components.}
\end{figure}

These observed characteristics suggest two possibilities:
\begin{itemize}
\item Wide, low-mass binaries are inhibited from forming by some characteristic
(disk size? disk mass?) of the star formation process.
\item Wide, low-mass binaries form, but are disrupted by tidal effects.
The binding energy of a binary system is linearly dependent on the total system mass
and inversely dependent on the separation, providing a natural explanation for 
the preference for low-q systems at large $\Delta$.
\end{itemize}
At present, we lack sufficient data to distinguish between these alternatives.
More extensive data on the prevalence of very close ($\Delta < 1$AU), spectroscopic
binaries and binary frequency in young, star-forming systems can shed more light
on this question.

\section {Summary and future prospects}

It is difficult to underestimate the progress made in very low-mass star/brown
dwarf research over the last few years. As recently as 1994 (Tinney's ESO conference),
questions were being raised over the very existence of brown dwarfs; now, we have
well over 50 confirmed candidates in the field, together with numerous examples in
open clusters (see Stauffer's talk, these proceedings). We can 
consider their detailed spectral energy distributions, and wrangle over the
appropriate means of botanising the sample. 

Larger numbers of brown dwarf binaries are being discovered, both in near-equal
mass systems and as companions to higher-mass main-sequence and evolved stars.
Some offer the
possibility of direct mass determination over reasonable timescales. The statistics of
the sample, particularly the semi-major axis and mass ratio distribution, will
provide insight into star formation mechanisms.

For the future, the brightest immediate prospects are offered by SIRTF. With
its high sensitivity at mid-infrared wavelengths, SIRTF will not only provide
much better data on the full energy distribution of ultracool dwarfs, improving
substantially bolometric magnitude determinations, but also should be capable of
detecting 300 to 400K (`room temperature') brown dwarfs. The last is particularly
important given the rapid evolutionary timescales for most brown dwarfs. The 
L dwarfs, and even T dwarfs, discovered to date by 2MASS, DENIS and SDSS are very much
the tip of a substantial iceberg, and all of our estimates of the mass function in the
field and the mass density contributed by substellar-mass objects are predicated on
reasonable, but untested, assumptions as to the scope of what is happening beneath
the surface. Direct observations wouldn't hurt.

\acknowledgements Thanks to the organisers for the invitation to speak at this 
conference, and thanks to imperial Gwen and svelte Daisy for brown dwarf modelling.
Much of the research described in this review was supported by a 
Core Project grant from 2MASS, and was carried out in collaboration with other members
of the Rare Objects team: Jim Liebert, Davy Kirkpatrick, John Gizis, Dave Monet, Conard
Dahn and Adam Burgasser.


\begin{references}

\reference Bailer-Jones, C.A.L., Mundt, R., 2000, ASP Conf. Ser. 198, 341 
(ed. R. Pallavicini, G. Micela \& S. Sciortinio)

\reference  Basri, G., Mohanty, S., Allard, F., Hauschildt, P.H., {\sl et al.}
 2000, ApJ, in press

\reference Burrows, A., Hubbard, W.B., Lunine, J.I. 1989  ApJ,  345, 939

\reference  Burrows, A., Marley, M., Hubbard, W.B., Lunine, J.I., {\sl et al.}
 1997, ApJ, 491, 856

\reference Burrows, A., Sharp, C.M. 1999, ApJ,  512, 843

\reference Dahn, C.C., Guetter, H.,  Harris, H., Henden, A., {\sl et al.} 1999, in
{\sl From Giant Planets to Cool Stars}, (ed. C. Griffiths \& M. Marley), 
ASP Conf. Proc. vol. 213, p. xxx

\reference  Delfosse, X., Tinney, C.G., Forveille, T., Epchtein, N., 
{\sl et al.} 1997, Astr. Astrophys., 327, L25

\reference  Duquennoy, A., Mayor, M., 1991, Astr. Astrophys., 248, 485

\reference  Fegley, B., Lodders, K. 1996, ApJ, 472, L37

\reference  Fischer, D.A., Marcy, G.W. 1992 ApJ, 396, 178

\reference Geballe, T., Kulkarni, S. R., Woodward, C. E., Sloan, G. C. 1996, ApJ,  467, L101

\reference Gizis, J.E., Monet, D., Reid, I.N., Kirkpatrick, J.D., Liebert, J., Williams, R.J. 2000,
AJ, in press

\reference  Kirkpatrick, J.D., Reid, I.N., Liebert, J., Cutri, R., 
{\sl et al.} 1999, Apj, 519, 802 (K99)

\reference  Kirkpatrick, J.D., Reid, I.N., Liebert, J., Gizis, J.E., {\sl et al.}
 2000, AJ, 120, 447 (K00)

\reference Koerner, D., Kirkpatrick, J.D., McElwain, M.W., 
Bonaventura, N.R.  1999, ApJL, 526, L25

\reference Krishnamurthi, A., Terndrup, D.M., Pinsonneault, M.H.,
Sellgren, K. {\sl et al.}, 1998, ApJ, 493, 914

\reference Leggett, S.K., Geballe, T. R., Fan, X., Schneider, D. P. {\sl et al.}, 
2000, ApJL, 536, L35

\reference Lodders, K. 1999, ApJ, 519, 793

\reference McLean, I.S., Wilcox, M.V., Becklin, E.E., Figer, D.F., {\sl et al.} 2000, 
ApJ, 533, L45

\reference Marcy, G.W., Butler, P. 1998, ARAA, 36, 57

\reference  Marley, M.S., Saumon, D., Guillot, T., Freedman, R.S., Hubbard, W.B.,
Burrows, A., Lunine, J.I. 1996, Science, 272, 1919

\reference Mart\'in, E.L., Brandner, W., Basri, G. 1999a, Science, 283, 1718

\reference Mart\'in, E.L., Delfosse, X., Basri, G., Goldman, N.,
{\sl et al.} 1999b, ApJ, 118, 2466

\reference  Mart\'in, E.L., Brandner, W., Bouvier, J., Luhman, K.L., 
{\sl et al.} 2000, ApJ, in press

\reference  Morgan, W.W., Keenan, P.C., Kellman, E. 1943, {\sl An Atlas of
Stellar Spectra}, (Chicago: Univ. of Chicago Press)

\reference Morgan, W.W. 1979, Ricerche Astronomiche, 9, 59 (IAU Coll. 47)

\reference Noll, K.S., Geballe, T.R., Leggett, S.K., Marley, M.S. 2000, ApJl, in press

\reference Oke, J. B., Cohen, J. G., Carr, M., Cromer, J., {\sl et al.} 1995,  PASP, 107, 375

\reference  Perryman, M.A.C., Brown, A.G.A., Lebreton, Y., G\'omez, A., 
{\sl et al.} 1998. A\&A, 331, 81

\reference Rebolo, R., Martin, E.L., Magazzu, A. 1992, ApJ, 389, L83

\reference Reid, I.N. 1992, MNRAS, 257, 257

\reference  Reid, I.N., Gizis, J.E., 1997a, AJ, 113, 2246

\reference  Reid, I.N., Gizis, J.E., 1997b, AJ, 114, 1992

\reference  Reid, I.N., Kirkpatrick, J.D., Liebert, J., Burrows, A., {\sl et al.} 1999,
ApJ,  521, 613

\reference Reid, I.N., Hawley, S.L. 2000, {\sl New Light on Dark Stars}, (Springer-Praxis: London, Heidelberg)

\reference Reid, I.N., Kirkpatrick, J.D., Gizis, J.E., Dahn, C.C., Monet, D.G.,
Williams, R.J., Liebert, J., Burgasser, A.J. 2000a, AJ, 119, 369

\reference Reid, I.N., Mahoney, S. 2000, MNRAS, in press

\reference Reid, I.N., Gizis, J.E., Kirkpatrick, J.D., Koerner, D.W. 2000b, AJ, in press

\reference Tinney, C.D., Tolley, A.J. 1999, MNRAS, 301, 1031

\reference Tokunaga, A., Kobeyashi, N. 1999, AJ, 117, 1010

\reference   Tsuji, T. 1964, Ann. Tokyo Obs. ser. II, 9, 1

\reference Tsuji, T., Ohnaka, K., Aoki, W. 1996, A\&A, 305, L1

\reference  Vogt, S.S., Allen, S.L., Bigelow, B.C., Bresee, L. 
{\sl et al.}, 1994, S.P.I.E., 2198, 362

\end{references}
\end{document}